\documentclass[aip,apl,preprint]{revtex4-1} 

\usepackage{graphicx}
\usepackage{textcomp}
\usepackage{hyperref}

\draft 

\begin{document}

\title{Combined metallic nano-rings and solid-immersion lenses for bright emission from single InAs/GaAs quantum dots} 

\author{Oliver Joe Trojak}
\affiliation{Department of Physics and Astronomy, University of Southampton, Southampton, SO17 1BJ, United Kingdom}

\author{Christopher Woodhead}
\affiliation{Physics Department, Lancaster University, Lancaster, LA1 4YB, United Kingdom}

\author{Suk-In Park}
\affiliation{Center for Opto-Electronic Materials and Devices Research, Korea Institute of Science and Technology, Seoul 136-791, South Korea}

\author{Jin Dong Song}
\affiliation{Center for Opto-Electronic Materials and Devices Research, Korea Institute of Science and Technology, Seoul 136-791, South Korea}

\author{Robert James Young}
\affiliation{Physics Department, Lancaster University, Lancaster, LA1 4YB, United Kingdom}

\author{Luca Sapienza}
\email{l.sapienza@soton.ac.uk}\homepage{www.quantum.soton.ac.uk}
\affiliation{Department of Physics and Astronomy, University of Southampton, Southampton, SO17 1BJ, United Kingdom}

\date{\today}

\begin{abstract}

Solid-state single-photon emitters are key components for integrated quantum photonic devices. However, they can suffer from poor extraction efficiencies, caused by the large refractive index contrast between the bulk material they are embedded in and air: this results in a small fraction (that can be as low as $<$1$\%$) of the emitted photons reaching free-space collection optics. To overcome this issue, we present a device that combines a metallic nano-ring, positioned on the sample surface and centered around the emitter, and an epoxy-based super-solid immersion lens, deterministically deposited above the ring devices. We show that the combined broadband lensing effect of the nano-ring and of the super-solid immersion lens significantly increases the extraction of light emitted by single InAs/GaAs quantum dots into free space: we observe enhancements in the emitted intensity by the nano-ring of up to a factor 20 and by the super-solid immersion lens of up to a factor 10. Such cumulative enhancements allow us to measure photon fluxes approaching 1 million counts per second, from a single InAs/GaAs quantum dot in bulk. The combined broad-band enhancement in the extraction of light can be implemented with any kind of classical and quantum solid-state emitter and opens the path to the realisation of scalable bright devices. The same approach can also be implemented to improve the absorption of light, for instance for small-area broadband photo-detectors.

\end{abstract}


\maketitle 

Improving the effiency in the extraction of light from solid-state emitters into free space is a challenge that often needs to be overcome when dealing with photonic devices. Total internal reflection due to the refractive index contrast between different materials results, in some cases, in even less than 1$\%$ of the emitted photons to be collected into free space. This is a clear drawback if one wants, for instance, to use light for quantum information processing, where high fluxes of single photons and minimised losses are required \cite{QIP}. Furthermore, the shape of the emitted light beam, as well as its angular divergence, are important parameters to control in order, for instance, to optimise the coupling efficiency of light into optical elements like objectives and single-mode fibers that can be used for propagating classical or quantum light over long distances with low loss. To improve the extraction efficiency of light, several photonic devices have been realised: optical cavities, which have the advantage of Purcell enhancements of the spontaneous emission rate \cite{QED}, achieving extraction efficiencies up to 65\,$\%$, although limited to relatively narrow bandwidths \cite{cavities}; nanowire geometries that allow guiding and shaping of the emitted light in a broad-band device, reaching efficiencies as high as 72\,$\%$\, \cite{nanowires}. Optical microlenses that provide broad-band improvements in the brightness of the emitted light of up to one order of magnitude are also widely used \cite{SILs} and have been deterministally fabricated around specific emitters \cite{Reitzenstein, Michler} or deposited from liquid phase \cite{Chris}. We have recently demonstrated that, by using an all-optical positioning technique, metallic nano-rings can be fabricated around specific emitters in order to obtain broad-band increases in the brightness of the emitted light, reaching enhancement factors as high as $\sim$20\, \cite{Oliver}. Here, we combine such metallic nano-rings with epoxy-based super-solid immersion lenses (super-SILs) \cite{Chris} to combine the broad-band enhancement effects of the two devices and obtain efficient emission from single InAs/GaAs quantum dots (QDs). We note that this technique is non-destructive and that both the ring and the super-SIL can be removed if required, leaving the properties of the emitters unchanged. Furthermore, such combined approach can be also implemented to improve the absorption of light in small-scale broadband photo-detectors.

We use a photoluminescence imaging technique, based on a double light-emitting diode (LED) illumination \cite{NComm} of a sample containing a low density of single InAs/GaAs QDs. A 455\,nm-LED is used to excite the QD emission and a 940\,nm-LED illuminates metallic markers that are deposited on the sample surface. An example of a photoluminescence image, collected with an electron-multiplying charge coupled device (EMCCD), of metallic alignment marks deposited on the sample surface and of the emission of single QDs, is shown in Fig.\,1a. By means of aligned electron-beam lithography, metal evaporation and chemical lift off, gold nano-rings are then deposited on the sample surface, centred around specific emitters (see Fig.\,1b - for more details, see Ref.\cite{Oliver}).
A super-SIL \cite{epoxy_SIL} is formed by dispensing liquid ultraviolet (UV)-curable epoxy (with refractive index of 1.54) onto a substrate over the region of the sample containing the metallic nanorings (see Fig.1c and 1d). The dispensing environment is filled with a liquid phase medium (glycerol), designed to increase the contact angle between the droplet edge and the substrate, by modification of the interfacial surface tensions \cite{liquid}, and, subsequently, the lens is solidified by exposure to UV light.

By carrying out micro-photoluminescence spectroscopy, we measure the intensity of light emitted by selected QDs at cryogenic temperatures ($\sim$10\,K), under non-resonant laser excitation (with a continuous-wave laser with emission wavelength of 780\,nm), as a function of pump power. We report the flux of photons on the first lens as a way to evaluate the emission brightness independently on the specific setup and detector used. We measure the transmission and detection efficiency of our setup and use them to extract the photon flux on the collection objective with a numerical aperture of 0.65. The results of our measurements are shown in Fig.\,2: the intensity of a single QD emission line is first enhanced by the metallic nano-ring and then further enhanced by the super-SIL. As shown, the effect is cumulative and allows us to detect photon fluxes as high as $\sim$900000 counts/s from a single QD in bulk. 
The enhancement factor deriving from the super-SIL depends on the relative position of the QD with respect to the apex of the lens: therefore, we observe a range of enhancements that depend on the position of the emitter. The statistics of the enhancement factors measured on QDs in bulk and within metallic nano-rings are shown in Fig.\,3. It is worth noting that different QD emission lines are enhanced by different factors by the metallic nano-rings, as already discussed in Ref.\,\cite{Oliver}, while the super-SIL effect is measured to be the same for the different emission lines of a given QD. Overall, we measure enhancement factors as high as $\times$20 by the nano-rings and up to $\times$10 by the super-SIL: for an ideally located super-SIL, cumulative enhancement factors as high as $\times$200 are therefore expected. Further enhancements in the collected intensity can be obtained by introducing a bottom mirror (like a metallic layer or a distributed Bragg reflector) that would reflect the fraction of the light that is currently emitted towards the substrate towards the collection optics.

The combined lensing effect that we have reported, that derives from the deterministically fabricated metallic nano-ring and the super-SIL, is wavelength insensitive in a wide range of infra-red wavelengths, therefore broad-band, and compatible with any kind of solid-state emitter of classical or quantum light on any substrate. Furthermore, given that the emitters are in bulk, detrimental effects, affecting for instance the coherence and/or stability of the light emission (that  derive from the proximity to surface states due to the presence of etched surfaces \cite{Jin} that are needed when fabricating, for instance pillar or photonic crystal structures) will be avoided in the emitters embedded within our devices.

In conclusion, we have shown how deterministally placed epoxy super-SILs can be implemented to enhance the extraction efficiency from single QDs and how such an effect can be combined to metallic nano-rings to realise broad-band devices allowing increases in the intensity of the light emitted by solid-state sources, potentially, as high as a factor 200. We have implemented such a combined system to show bright emission from single QDs in bulk, reaching photon fluxes approaching 1 million counts per second, in a scalable and easy to fabricate device: the positioning of single QDs can be carried out in an automated system \cite{Jin_positioning}, the dimensions of the rings are compatible with photolithography and nano-imprint, and the super-SIL deposition requires only optical microscopy tools.

\section*{Acknowledgments}

LS acknowledges support by the Royal Society through the grant RG130684.
RJY acknowledges support by the Royal Society through a University Research Fellowship (UF110555 and UF160721) and support by the Air Force Office of Scientific Research (FA9550-16-1-0276) and the Engineering and Physical Sciences Research Council (EP/K50421X/1 and EP/L01548X/1). SIP and JDS acknowledge the support from the KIST institutional flag-ship program.

\begin{figure}[htbp!]
\centering
\includegraphics[width=0.7\linewidth]{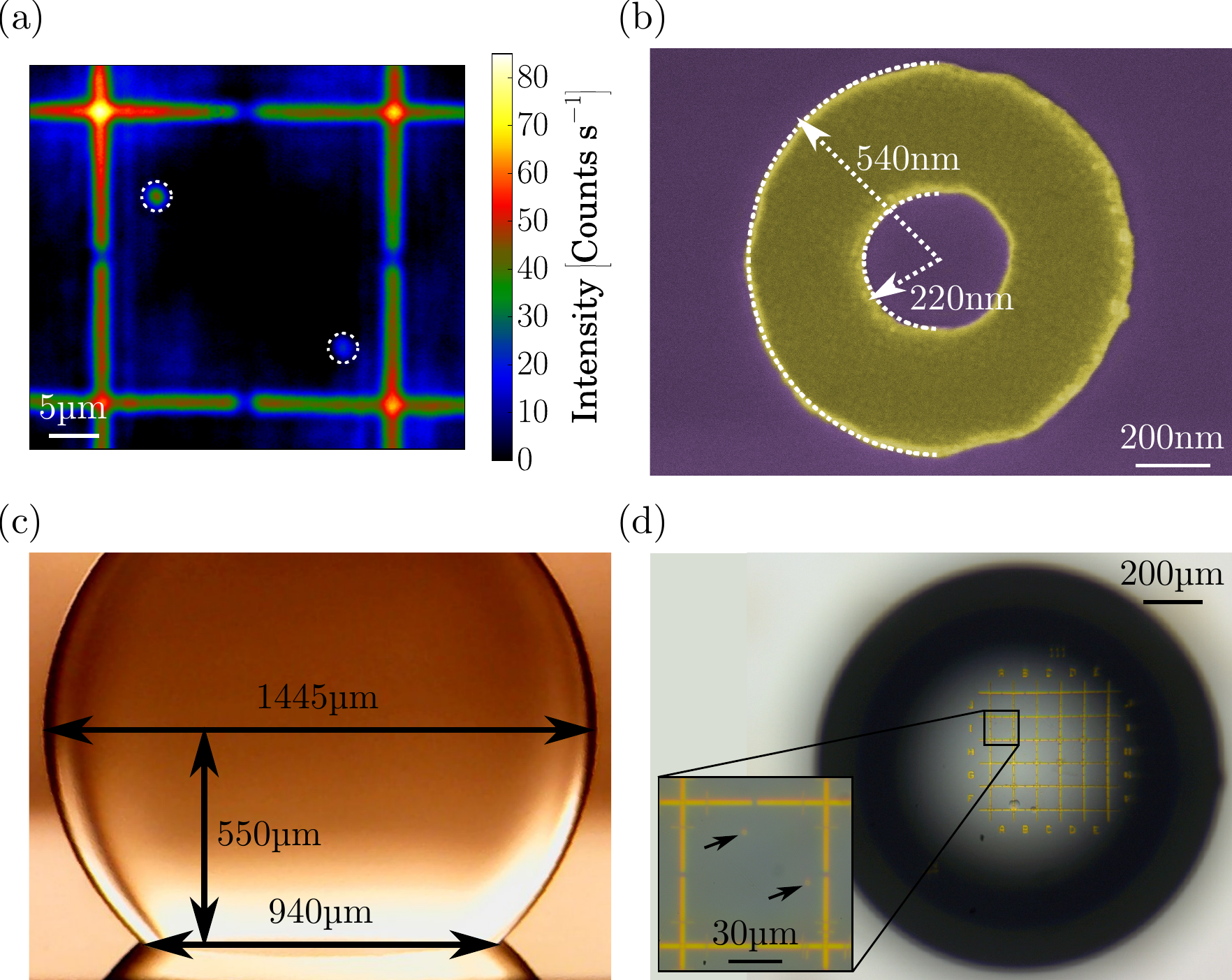}
\caption{(a) Example of a photoluminescence image, obtained under double LED illumination on an EMCCD camera (see main text for more details), showing metal alignment markers and the photoluminescence emission from two QDs, highlighted by the dashed lines (before super-SIL deposition, EMCCD settings: 2\,s integration, 100 acquisitions). (b) False colour scanning electron micrograph of a metallic nano-ring, with critical dimensions shown. (c) Side profile of a super-SIL with critical dimensions shown. (d) Optical micrograph of a super-SIL deposited on a sample in correspondence to metallic alignment markers and nano-rings. Inset: Zoom-in of a single field showing metallic alignment markers and nano-rings positioned around selected QDs (highlighted by the arrows).}
\end{figure}

\begin{figure}[h!]
\centering
\includegraphics[width=0.8\linewidth]{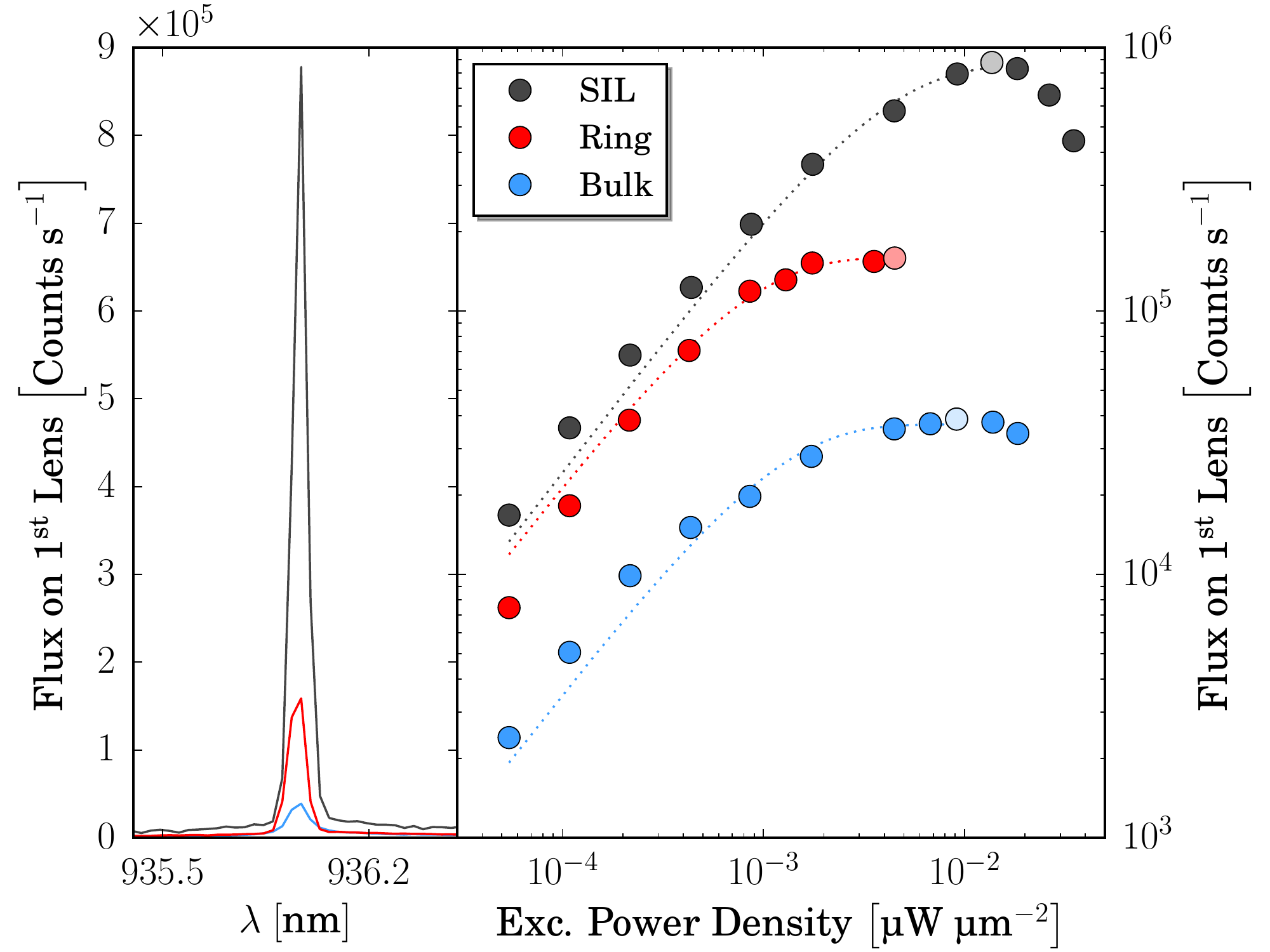}
\caption{Left: Photoluminescence spectra collected at saturation pump power for the emitter in bulk (blue line), after ring deposition (red line) and after super-SIL placement (black line).  Right: Intensity of the emission lines shown in panel (a) plotted as a function of laser excitation power density  (lighter dots correspond to the peak intensity measured from the spectra shown in the left panel, dashed lines are fits to the data).}
\end{figure}

\begin{figure}[h!]
\centering
\includegraphics[width=0.8\linewidth]{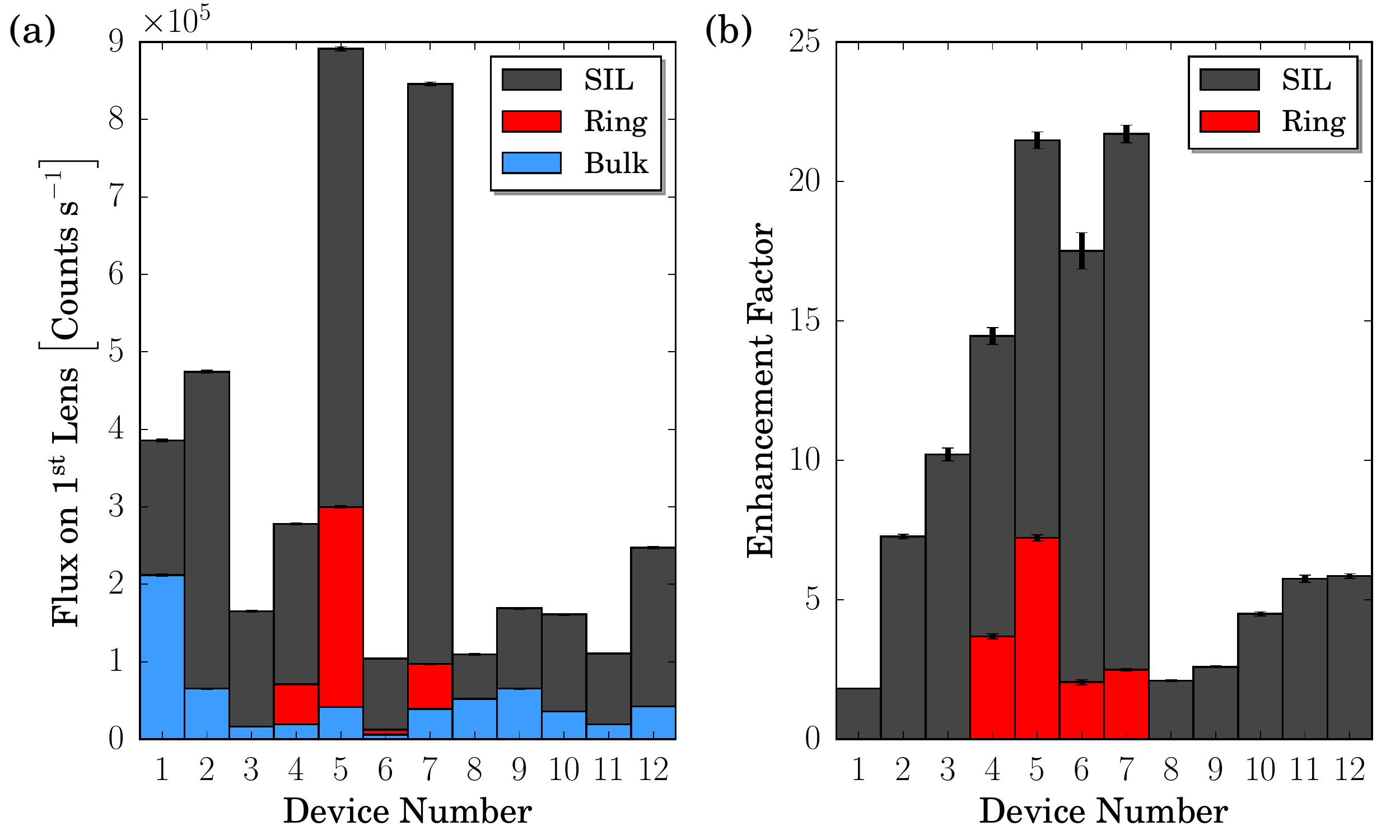}
\caption{Left panel: Histograms showing the photon flux on the first lens placed above the QD sample, at saturation pump powers, for the emitters in bulk (blue), with a nano-ring deposited on the sample surface (red), with a super-SIL and/or nano-ring deposited on the sample surface (black). Right panel: Enhancement factors deriving from the nano-rings and super-SILs, measured on the QDs shown in the left panel, using the same colour coding. (Note that for devices number 4 to 7, the black bars represent the cumulative enhancement factor of the nano-ring and of the super-SIL).}
\end{figure}


\begin{thebibliography}{50}

\bibitem{QIP} J.L. O’Brien, A. Furusawa, J. Vuckovic, Nature Photonics \textbf{3}, 687 (2009).

\bibitem{QED} M. Pelton, Nature Photonics \textbf{9}, 427 (2015).

\bibitem{cavities} O. Gazzano and G.S. Solomon , J. Opt. Soc. Am. B \textbf{33}, C160 (2016).

\bibitem{nanowires} J. Claudon, J. Bleuse, N. S. Malik, M. Bazin, P. Jaffrennou, N. Gregersen, C. Sauvan, P. Lalanne, J. Gerard, Nat. Photonics \textbf{4}, 174 (2010). 

\bibitem{SILs} K.A. Serrels, E. Ramsay, P.A. Dalgarno, B.D. Gerardot, J.A. O'Connor, R.H. Hadfield, R.J. Warburton,  D.T. Reid, J. Nanophotonics \textbf{2}, 021854 (2008). 

\bibitem{Reitzenstein} M. Gschrey et al., Nature Communications \textbf{6}, 7662 (2015).

\bibitem{Michler} M. Sartison, S.L. Portalupi, T. Gissibl, M. Jetter, H. Giessen, P. Michler, Sci. Rep. \textbf{7}, 39916 (2017). 

\bibitem{Chris} C.S. Woodhead, J. Roberts, Y.J. Noori, Y. Cao, R. Bernardo-Gavito, P. Tovee, A. Kozikov, K. Novoselov, R.J. Young, 2D Materials \textbf{4}, 015032 (2017).

\bibitem{Oliver} O.J. Trojak, S.I. Park, J.D. Song, L. Sapienza, Applied Physics Letters \textbf{111}, 021109 (2017).

\bibitem{NComm} L. Sapienza, M. Davanco, A. Badolato, K. Srinivasan, Nat. Commun. \textbf{6}, 7833 (2015).

\bibitem{epoxy_SIL} B. Born, E.L. Landry, J.F. Holzman, IEEE Photonics J. \textbf{2}, 873 (2010).


\bibitem{liquid} C.C. Lee, S.Y. Hsiao, W. Fang, Int. Solid-State Sensors, Actuators and Microsystems 2086 (2009).

\bibitem{Jin} J. Liu et al., arXiv:1710.09667 (2017).

\bibitem{Jin_positioning} J. Liu, M. Davanco, L. Sapienza, K. Konthasinghe, J.V. De Miranda Cardoso, J.D. Song, A. Badolato, K. Srinivasan, Review of Scientific Instruments \textbf{88}, 023116 (2017).




\end{thebibliography}
\end{document}